\documentclass[prb,twocolumn,showpacs,preprintnumbers,amsmath,amssymb, superscriptaddress]{revtex4}

\usepackage{graphicx}
\usepackage{graphics}
\usepackage{dcolumn}
\usepackage{bm}
\usepackage{subfigure}

\begin{document}

%\preprint{APS/123-QED}

\title{Effect of Chemical Pressure on the Charge Density Wave Transition in Rare-earth Tritellurides $R$Te$_3$}

\author{N. Ru}
\affiliation{Geballe Laboratory for Advanced Materials and Dept. of Applied Physics, Stanford University, Stanford, CA 94305 (USA)\\
}

\author{C. L. Condron}
    \affiliation{Stanford Synchrotron Radiation Laboratory, Stanford Linear Accelerator Center, Menlo Park, CA 94025 (USA)}

\author{G. Y. Margulis}
\author{K. Y. Shin}
\affiliation{Geballe Laboratory for Advanced Materials and Dept. of Applied Physics, Stanford University, Stanford, CA 94305 (USA)\\
}

\author{J. Laverock}
\author{S. B. Dugdale}

\affiliation{H. H. Wills Physics Laboratory, University of Bristol, Tyndall Avenue, Bristol BS8 1TL (UK)}

\author{M. F. Toney}
    \affiliation{Stanford Synchrotron Radiation Laboratory, Stanford Linear Accelerator Center, Menlo Park, CA 94025 (USA)}

\author{I. R. Fisher}
    \affiliation{Geballe Laboratory for Advanced Materials and Dept. of Applied Physics, Stanford University, Stanford, CA 94305 (USA)\\
}

\date{\today}

\begin{abstract}

The charge density wave transition is investigated in the bi-layer family of rare earth tritelluride
$R$Te$_3$ compounds ($R$ = Sm, Gd, Tb, Dy, Ho, Er, Tm) via high resolution x-ray diffraction
and electrical resistivity. The transition temperature increases monotonically with increasing lattice parameter
 from 244(3) K for TmTe$_3$ to 416(3) K for SmTe$_3$.
The heaviest members of the series, $R$ = Dy, Ho, Er, Tm, are observed to have a second transition at a lower temperature, which marks the onset of an additional CDW with wavevector almost equal in magnitude to the first, but oriented in the perpendicular direction.
\end{abstract}

\pacs{71.45.Lr, 61.44.Fw, 61.10.Nz, 72.15.-v}

\maketitle

\section{Introduction}
Charge density waves (CDWs) are electronic instabilities found in
low-dimensional materials with highly anisotropic electronic structures.\cite{gruner1}
Since the CDW is predominantly driven by Fermi-surface (FS) nesting, it is especially sensitive to pressure-induced changes
in the electronic structure. A well known example is NbSe$_2$,
for which the CDW can be completely suppressed by an applied pressure
of 35 kbar, favoring the competing superconducting phase.\cite{NbSe2}  Chemical pressure (the incorporation of larger or smaller ions to expand or contract the crystal lattice)
can be used to mimic the effect of external pressure,
providing a valuable tuning parameter for such materials.
In this regard, rare-earth containing compounds are particularly valuable because
the lattice parameter can be varied over a wide range in an almost continuous
fashion while keeping the band filling essentially unchanged.
The rare earth tritelluride $R$Te$_3$ compounds form for almost the entire rare-earth series,
with $R$ =  La-Nd, Sm, Gd-Tm, \cite{bucher, norling, dimasi_1995}
and provide a unique opportunity to follow the effect of chemical pressure on FS nesting and CDW formation.

The rare-earth tritellurides have a crystal structure that is layered and weakly orthorhombic
(space group Cmcm), \cite{norling} consisting of double layers of nominally square-planar Te sheets,
separated by corrugated $R$Te slabs.
In this space group, the long $b$-axis is perpendicular to the Te planes.
For most of the rare earths, the material has an incommensurate lattice modulation at room temperature,
with a single in-plane wave-vector of approximately 2/7 $c^*$ ($c^*$ = 2$\pi/c$). \cite{dimasi_1995, malliakas_2005, iyeiri, kim, fang}
For the heaviest rare earths, the same modulation has been observed
slightly below room temperature.\cite{malliakas_2006}
The band structure has been calculated for the unmodulated structure, yielding a simple FS
consisting of slightly warped inner and outer diamond sheets formed
from $p_x$ and $p_z$ orbitals of Te atoms in the square planar layers, both doubled due to the effects of bilayer splitting,
with minimal $b$-axis dispersion.\cite{laverock}
The susceptibility $\chi(q)$ has a peak at a wave vector
which is identical, to within experimental error,
with the observed incommensurate lattice modulation. \cite{yao, mazin_comm, chiq}
Angle resolved photoemission spectroscopy (ARPES) has revealed that
portions of the FS which are nested by this wave vector are
indeed gapped, implying that the gain in one-electron energy contributes to the CDW formation. \cite{gweon, brouet, komoda, maxgap}
There is a small but finite electronic contribution to the heat capacity,
and the resistivity is observed to be metallic down to the lowest temperatures,
confirming the presence of a small fraction of ungapped, reconstructed FS in the CDW state. \cite{dimasi_1994, iyeiri, ru_2006}

ARPES measurements have shown that the maximum value of the CDW gap
scales with the lattice parameter, with published values of $\sim$280 meV for SmTe$_3$ \cite{gweon}
and $\sim$400 meV for CeTe$_3$. \cite{brouet}
This trend has been confirmed by optical conductivity measurements,
which also reveal that the remaining fraction of ungapped FS in the CDW state is larger for
compounds with smaller lattice parameters. \cite{sacchetti, sacchetti_pressure}
However, to date the most prominent feature of this class of material, the CDW transition,
has not been reported for this family of compounds.
In this paper we identify the CDW transition via high resolution x-ray diffraction and resistivity measurements,
and show that the transition temperature $T_{c}$ varies by a remarkable amount, over 200 K, across the rare earth series.
In addition, we observe a second transition for the heaviest members of the series - $R$ = Dy, Ho, Er, Tm - and we show for ErTe$_3$ that this marks the onset of a second CDW with a wavevector $q_2$ $\approx$ 1/3 $a^*$ of similar magnitude and perpendicular to the first.  The presence of this second transition is intimately linked to the effect of chemical pressure on the electronic structure.

\section{Experimental Methods}%%%%%%%%%%%%%%%%%%%%%%%%%%%%%%%%%%%%%%%%%%%%%%%%%%%%%%%%%%%%%%%%%%%%%%%%%%%%%%%%%%%%%%%

Single crystals of $R$Te$_3$ were grown by slow cooling a binary melt as described previously.\cite{ru_2006}

The electrical resistivity was measured up to 450 K for geometric bars
cut and cleaved from the larger as-grown crystals, using
an LR-700 AC resistance bridge operating at 16 Hz. Typical contact resistances of
1 - 3 $\Omega$ ­ were achieved using sputtered gold pads. In-plane
measurements were made for arbitrary current directions
in the ac plane, using a standard 4 point contact
geometry. The b-axis resistivity  was measured using
a modified Montgomery geometry, with one current and
one voltage contact on the top face of the plate-like crystal,
and the other voltage and current contacts on the
bottom face. The aspect ratio of the equivalent isotropic
crystal\cite{mont, logan} was typically 1:1 or slightly less, indicating
that this technique may slightly underestimate the absolute
value of the b-axis resistivity. Several measurements
were made for each compound in the series.

High resolution single crystal x-ray diffraction experiments were performed at the Stanford Synchrotron Radiation Laboratory (SSRL) on Beamlines 7-2 and 11-3.
Data were collected in reflection geometry using x-ray energies of 9 to 13 keV.  Either 1 or 2 milliradian slits or a Ge(111) crystal analyzer were used to define the diffracted beam.
The sample was mounted in an Anton-Paar furnace for heating above room temperature,
and kept in a flow of helium gas to prevent oxidation.  For measurements below room temperature, an ARS displex refrigerator was used, with the sample under vacuum.  Uncertainties in absolute values of both the CDW wavevector and superlattice peak intensities were larger for experiments performed in the displex refrigerator.  Experiments were performed on TbTe$_3$ and ErTe$_3$ as representative examples of compounds in the series with one and two transitions respectively.

\section{Results}%%%%%%%%%%%%%%%%%%%%%%%%%%%%%%%%%%%%%%%%%%%%%%%%%%%%%%%%%%%%%%%%%%%%%%%%%%%%%%%%%%%%%%%%%%%%%%%%%%%%%

\subsection{Resistivity} %%%%%%%%%%%%%%%%%%%%%%%%%%%%%%%%%%%%%%%%%%%%%%%%%%%%%%
\begin{figure}[ht]
\includegraphics[width=3.5in]{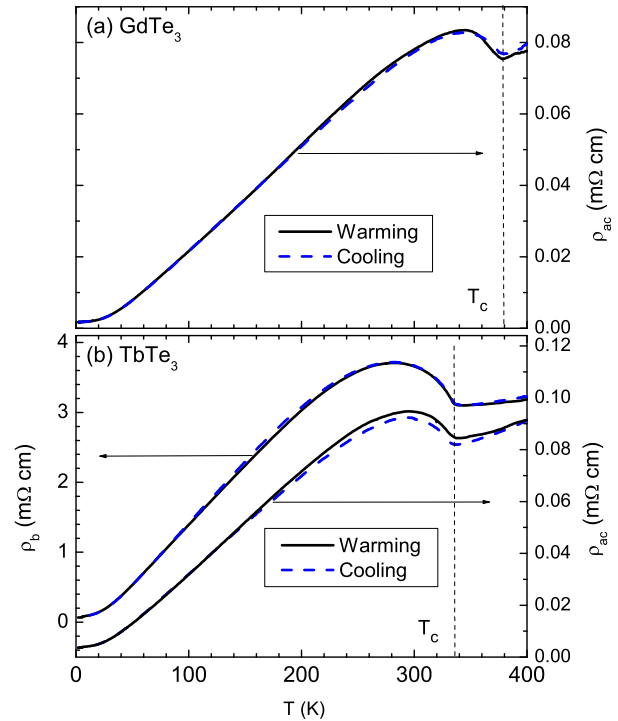}
\caption{\label{fig:GdTb} Temperature dependence of the resistivity of GdTe$_3$(a) and  TbTe$_3$(b).
Data are shown for currents oriented along the $b$-axis ($\rho_{b}$, left axis) and arbitrary in-plane orientations ($\rho_{ac}$, right axis).  Vertical axes are offset for clarity.
Both warming (solid lines) and cooling (dashed lines) cycles are shown.
$T_{c}$ is marked by a vertical line. }
\end{figure}

Representative resistivity data for GdTe$_3$ and TbTe$_3$ are shown in Fig.~\ref{fig:GdTb}(a) and (b), respectively.
Absolute values of the resistivity have been normalized to the average of several measurements of different samples.
As has been previously observed for this family of compounds,  the resistivity is strongly anisotropic.\cite{dimasi_1994, ru_2006}  For both compounds, a clear feature can be seen for both current orientations, with an onset at $T_{c}$ = 377(3) and 336(3) K for GdTe$_3$ and TbTe$_3$ respectively.  There is no hysteresis between warming and cooling cycles.  As described in the next section, x-ray diffraction measurements show that this feature marks the CDW transition for TbTe$_3$.
Excepting the loss of spin-disorder scattering at $T_N$ = 5.6 K for TbTe$_3$,
no other features were observed in either the resistivity or its derivative,
suggesting that there are no additional CDW phase transitions below $T_{c}$ for either compound.  The resistivity curves for SmTe$_3$ are similar, with $T_c$ = 416(3) K.  Data for the lightest members of the series, LaTe$_3$, CeTe$_3$, NdTe$_3$, and PrTe$_3$, are linear up to our maximum temperature and these compounds are presumed to have CDW transition temperatures greater than 450 K.

\begin{figure*}[ht]
\includegraphics[width=7in]{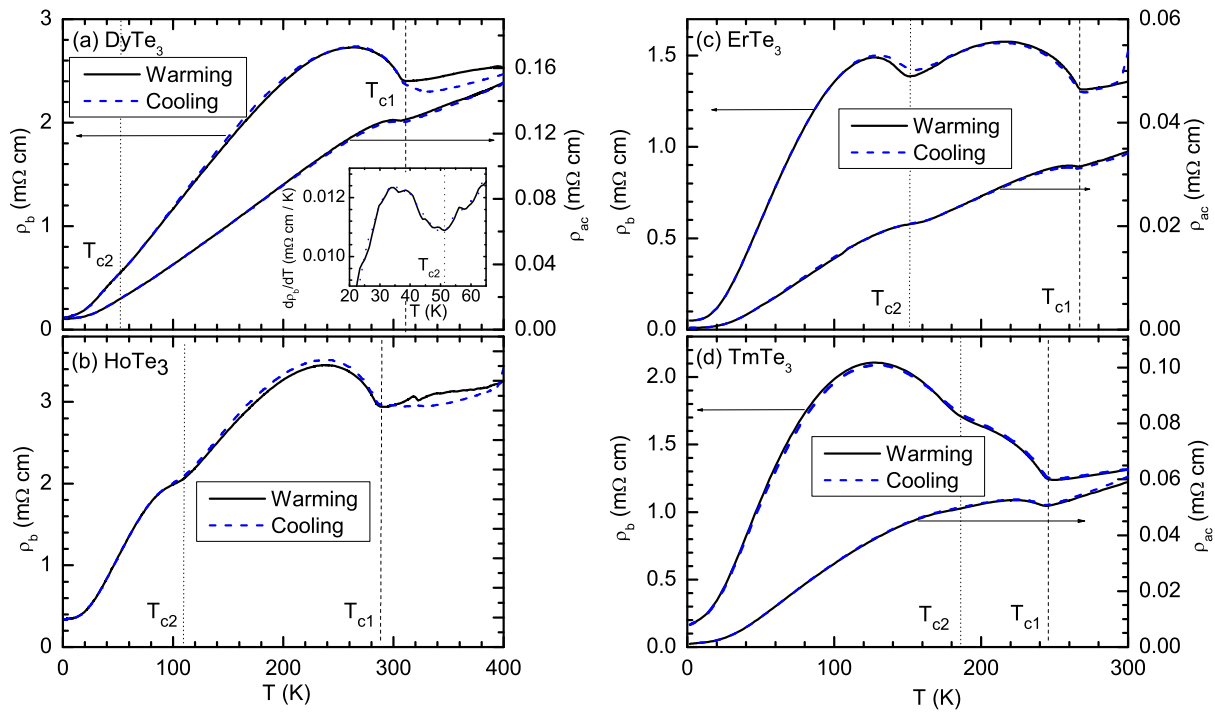}
\caption{\label{fig:DyHoErTm} Temperature dependence of the resistivity of (a) DyTe$_3$ (Inset: d$\rho$/dT), (b) HoTe$_3$, (c) ErTe$_3$ and  (d) TmTe$_3$.
Data are shown for currents oriented along the $b$-axis ($\rho_{b}$, left axis) and arbitrary in-plane orientations ($\rho_{ac}$, right axis).
Both warming (solid lines) and cooling (dashed lines) cycles are shown.
Both CDW transitions are marked for each compound.}
\end{figure*}

The resistivity data for the heavier members of the series ($R$ = Dy, Ho, Er, and Tm) are shown in Fig.~\ref{fig:DyHoErTm} (a) - (d). These are qualitatively different than those of the lighter rare earths, showing two transitions at $T_{c1}$ and $T_{c2}$.  Again, data are shown for both heating and cooling cycles, as well as for in-plane and b-axis orientations.  Looking at the b-axis data - for which the current direction is better defined - the feature at $T_{c2}$ is largest for the heaviest member of the series, TmTe$_3$.  This feature then decreases in both magnitude and temperature for the lighter rare earths, vanishing for compounds lighter than DyTe$_3$.\cite{TbTe3_surface} The derivative d$\rho$/dT for DyTe$_3$ is shown in the inset to Fig.~\ref{fig:DyHoErTm}(a) to highlight the feature at $T_{c2}$.  In the next section we identify for ErTe$_3$ that $T_{c2}$ marks the onset of a CDW with wavevector perpendicular to the first.

\begin{figure}
\includegraphics[width=3.5in]{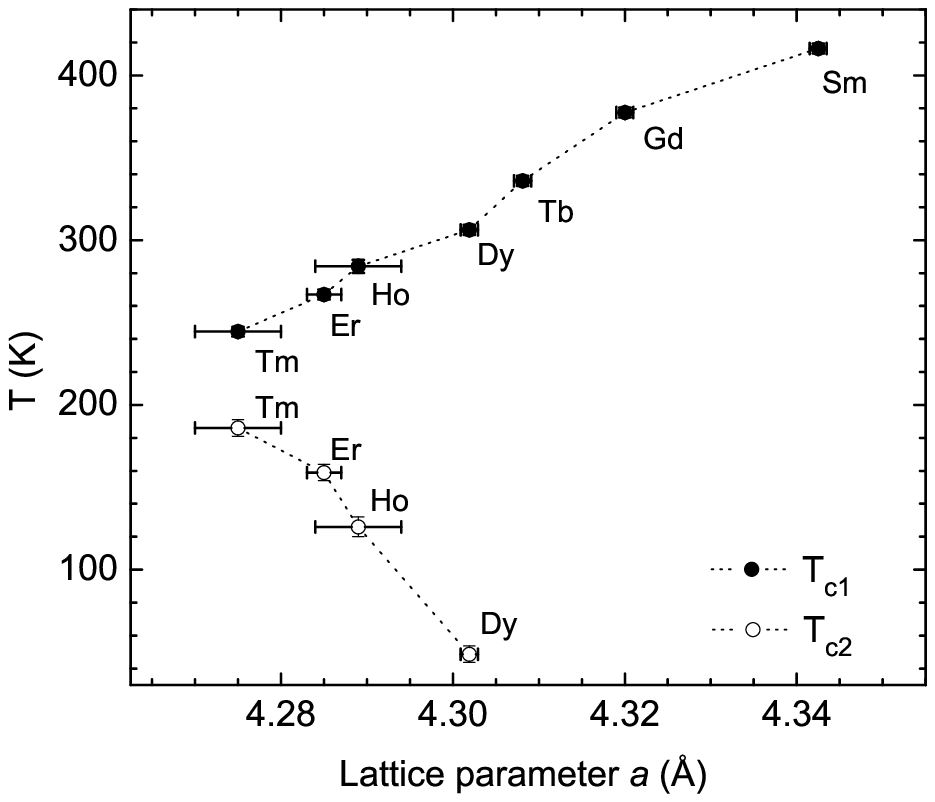}
\caption{\label{fig:TCDW} $T_{c1}$ and $T_{c2}$, as obtained from resistivity measurements, plotted as a function of in-plane lattice parameter $a$ at 300 K \cite{lattice}
for several compounds in the $R$Te$_3$ series (labeled).
Note that the lattice parameter is smaller for the heavier members of the $R$Te$_3$ series.
Dashed lines are drawn between points to guide the eye. }
\end{figure}

Figure~\ref{fig:TCDW} shows both CDW transition temperatures plotted as a function of the in-plane a-axis lattice parameter, \cite{lattice} which provides a measure of the chemical pressure.  The first transition temperature $T_{c1}$ ranges from a low of 244(3) K for TmTe$_3$, the compound with the smallest lattice parameter in the series, and increases monotonically with increasing lattice parameter (decreasing chemical pressure).  In contrast, the second transition temperature $T_{c2}$ shows the opposite trend, \textit{decreasing} as the lattice parameter increases, eventually vanishing halfway across the series.

\subsection{X-ray diffraction of TbTe$_3$}

\begin{figure}
\includegraphics[width=3.5in]{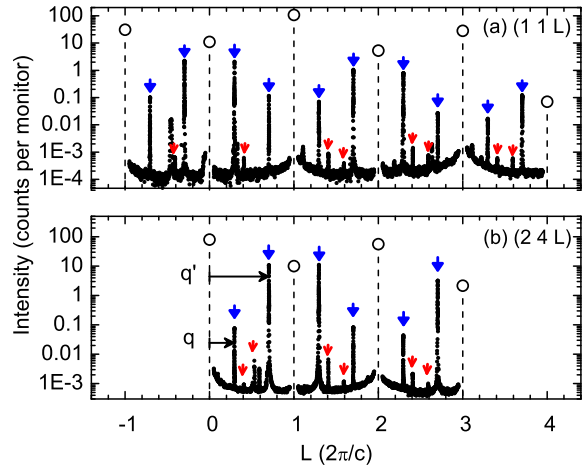}
\caption{\label{fig:LongKscans}Representative x-ray diffraction patterns for TbTe$_3$ along (a) (1 1 L) and (b) (2 4 L) at room temperature.  Intensities of Bragg peaks from the average structure are indicated by open circles and are scaled down by a factor of 1000.  First and second harmonics of the CDW modulation are marked by blue and red arrows, respectively (color online).  The wavevector $q$ and the alternate wavevector $q'$ are marked. }
\end{figure}

X-ray diffraction data for TbTe$_3$ at room temperature reveal an incommensurate modulation
wavevector $q$ = 0.296(4) $c^*$ and its harmonic 2$q$.  Here, for clarity, we reference the superlattice peak inside the first Brillouin zone, resulting in a wavevector $q$ $\approx$ 2/7 $c^*$, although ARPES\cite{brouet} and STM\cite{fang} results suggest that the equivalent $q'$ = $c^*$-$q$ = 0.704(4) $c^*$ $\sim$ 5/7 $c^*$ has more physical meaning.    Figure~\ref{fig:LongKscans} shows representative diffraction patterns along the direction of the CDW wavevector for TbTe$_3$, as measured at SSRL Beamline 11-3 using 2 milliradian slits.  Intensities of Bragg peaks from the average crystal structure are indicated by open circles and are scaled down by a factor of 1000.  First and second harmonic superlattice peaks are marked by blue and red arrows (color online).  Higher harmonics were not observed.  The wavevector $q$ and the equivalent $q'$ are indicated on the lower panel.  Both the average structure Bragg peaks and the CDW superlattice peaks alternate in intensity, which can be attributed to the presence of two inequivalent Te atoms in the Te planes.\cite{norling}

\begin{figure}
\includegraphics[width=3.7in]{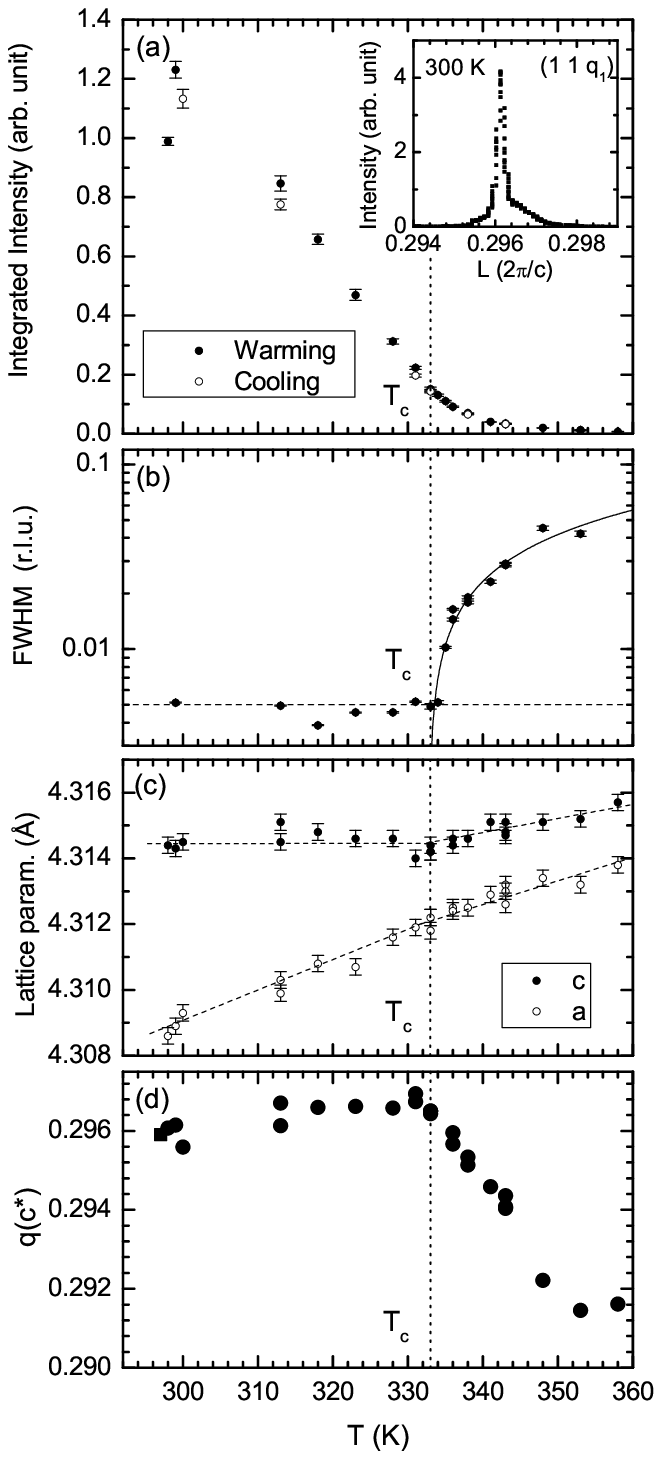}
\caption{\label{fig:Tbtransition} (a) Temperature dependence of the integrated intensity of the
(1 1 $q$) superlattice peak of TbTe$_3$ through $T_c$ for increasing and decreasing temperatures. Inset: Raw data showing L scan for (1 1 q) at room temperature.
(b) FWHM in the out of plane direction for the (1 7 $q$) superlattice peak. A dashed horizontal line marks the limit of resolution.  The solid curve is $\xi^{-1} \sim (T-T_{c})^{\gamma}$ with $T_{c}$= 332.8(5) K and $\gamma = 2/3$.
(c) The in-plane lattice parameters $a$ and $c$, with lines drawn to guide the eye.  (d) The absolute value of $q$ as a function of temperature. $T_{c}$ is indicated by a dashed vertical line in all four panels.}
\end{figure}

The superstructure peaks in TbTe$_3$ are very sharp and, along with the average lattice peaks, were resolution-limited in our experiments.  The inset to Fig.~\ref{fig:Tbtransition}(a) shows a typical superlattice peak as measured in high-resolution mode, using a Ge(111) crystal analyzer at SSRL Beamline 7-2.  From this we can draw lower bounds on the CDW correlation length $\xi$, calculated to be 1.8 $\mu$m within the Te planes and 0.5 $\mu$m perpendicular to the planes.

Heating measurements were taken with the sample mounted in an Anton-Paar furnace at SSRL Beamline 7-2 using a Ge(111) crystal analyzer.  The integrated intensity of the (1 1 $q$) peak in TbTe$_3$ was followed as a function of temperature.  Fig.~\ref{fig:Tbtransition}(a) shows the integrated intensity through the CDW transition for increasing and decreasing temperatures.  The intensity rapidly decreases on heating from room temperature with no observable hysteresis, indicative of a second-order CDW transition as described theoretically.\cite{yao}  Some scattering intensity, attributed to fluctuations, was observed above $T_{c}$ up to 363 K, beyond which the peaks were too weak to distinguish from background. Error bars shown are obtained from Lorentz fits to the individual scans.  The full width at half maximum (FWHM, Fig.~\ref{fig:Tbtransition}(b)) is resolution limited at lower temperatures and increases sharply at the CDW transition.  The FWHM, being inversely proportional to the correlation length $\xi$,  is fit in the interval from 334 to 353 K to the expression $\xi^{-1} \approx (T-T_{c})^{\gamma}$, where $\gamma$ is taken as $2/3$ to correspond to the X-Y model universality class (complex order parameter and three dimensions).  From this we obtain a transition temperature $T_c$ = 332.8(5) K, which is drawn as a dotted vertical line through all four panels of Fig.~\ref{fig:Tbtransition}.  Within error, this value of $T_c$ corresponds with that obtained from resistivity measurements (336(3) K).

\begin{figure}
\includegraphics[width=3.5in]{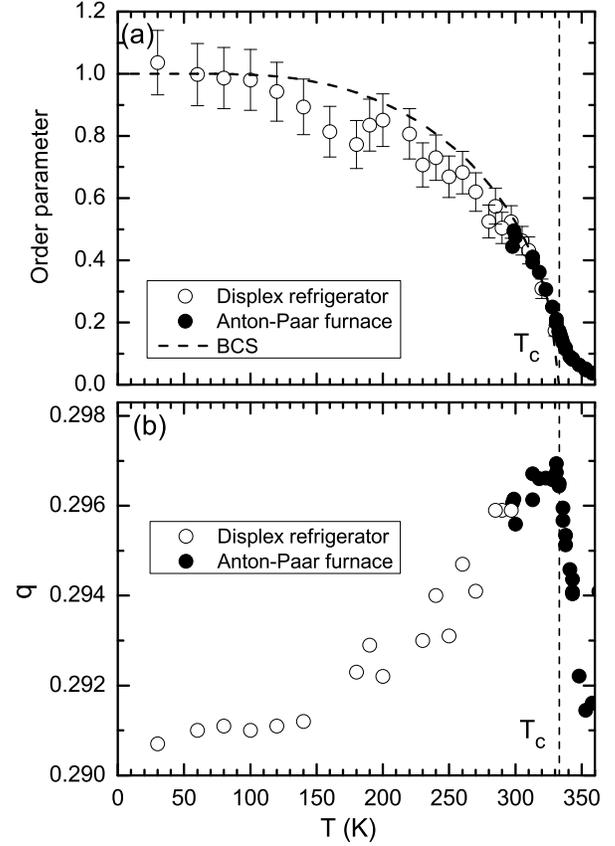}
\caption{\label{fig:TbOrderparameter} (a) Temperature dependence of the square root of the integrated intensity of the
(1 1 $q$) superlattice peak of TbTe$_3$, which is proportional to the order parameter.  Data are normalized to unity at T = 0.  (b) The absolute value of $q$ over the same temperature range. $T_c$ is indicated by a dashed vertical line in both panels.}
\end{figure}

The CDW transition in TbTe$_3$ is also reflected in the temperature dependence of the in-plane lattice parameters $a$ and $c$ (Fig.~\ref{fig:Tbtransition}(c)). Above $T_{c}$, $a$ and $c$ have a similar dependence, increasing linearly with temperature, and with a small relative difference of only 0.031$\%$ at 370K. Formation of the CDW appears to ``stretch'' the lattice from its expected value along the direction of the modulation wave vector ($c$ axis), whilst slightly compressing the lattice in the perpendicular direction, such that the relative difference between $a$ and $c$ at 300 K, 0.13$\%$, is larger than above the CDW transition.

The absolute value of the CDW wavevector for TbTe$_3$ is plotted in Fig.~\ref{fig:Tbtransition}(d).  The absolute value of $q$ does not vary significantly with temperature between room temperature and $T_c$.  For temperatures above $T_c$, the absolute value of $q$ decreases with increasing temperature.

Measurements for TbTe$_3$ were extended from room temperature down to 25 K by using an ARS displex refrigerator at SSRL Beamline 7-2 using 1 milliradian slits.   Fig.~\ref{fig:TbOrderparameter}(a) shows the temperature dependence of the square root of the integrated intensity, which is proportional to the order parameter,\cite{gruner1} together with the standard BCS curve.  Data have been normalized by taking the average of the first four data points, from 30 to 100 K, and setting this value to unity.  The close correspondence between the data and the BCS result confirm that the system can be treated within the weak-coupling limit.  Error bars shown are $\pm 20\%$ of the integrated intensity, and represent the scatter associated with difficulties of centering the sample within the refrigerator.  The apparent dip in the experimental data at approximately 150 K does not correlate with any additional features in the resistivity, and is most likely an experimental artifact.

Finally, the wavevector $q$ (Fig.~\ref{fig:TbOrderparameter}(b))remains incommensurate to the lowest temperatures, and does not exhibit a lock-in transition.

\subsection{X-ray diffraction of ErTe$_3$}

 To explore the case with two transitions, single crystal x-ray diffraction experiments were performed for ErTe$_3$ with sample mounted in a ARS displex refrigerator at SSRL Beamline 7-2.  An incommensurate modulation similar to TbTe$_3$ with $q_1$ $\approx$ 2/7 $c^*$ was found in the temperature region $T_{c2} < T < T_{c1}$, as also noted by Malliakas et al.\cite{malliakas_2006}  For temperatures $T < T_{c2}$, an additional ordering along $a^*$ was observed, perpendicular to the first wavevector. As investigated at 10K using a Ge(111) crystal analyzer, the wavevector of this additional ordering was $q_2$ = 0.343(5) $a^*$ whereas that of the original ordering was $q_1$ = 0.300(5) $c^*$.  Again, these values are cited in the first Brillouin zone even though the equivalent values $q_1'$ = $c^*$ - $q_1$  = 0.700(5) $c^*$ and $q_2'$ = $a^*$ - $q_2$ = 0.657(5) $a^*$,  may have more physical meaning for this system. \cite{brouet, fang}  The absolute value of $q_2$ is close but not identical to that of $q_1$.
 
\begin{figure}
\includegraphics[width=3.5in]{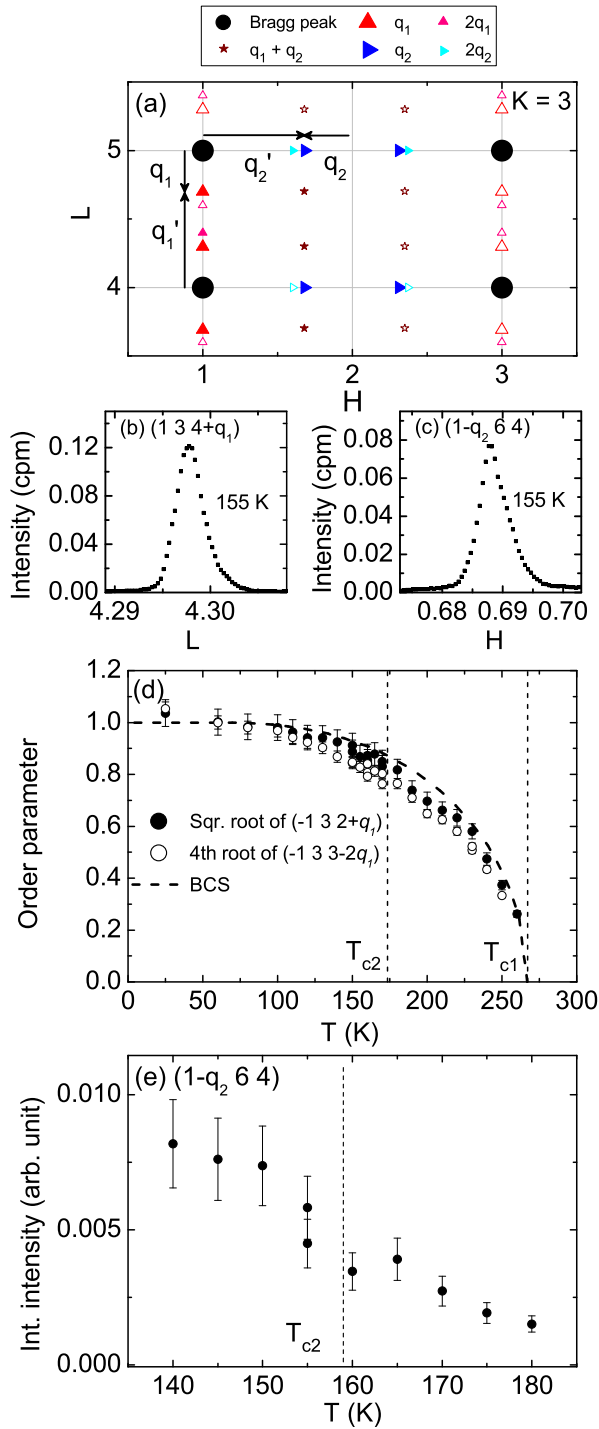}
\caption{\label{fig:ErXRD} X-ray diffraction of ErTe$_3$. (a) Reciprocal lattice map.  Solid symbols mark diffraction peaks found at 10 K.  Hollow symbols mark peaks assumed to be present from symmetry. (color online) (b) L scan for (1 3 4+$q_1$) at 155 K. (c) H scan for (1-$q_2$ 6 4) at 155 K.  (d) The BCS order parameter (dashed line), plotted with the square root of the integrated intensity of (-1 3 2+$q_1$), and the fourth root of the integrated intensity of (-1 3 3-2$q_1$), both of which are normalized to unity at T = 0.  (e) Temperature dependence of the integrated intensity of (1-$q_2$ 6 4).}
\end{figure}

A section of reciprocal space explored at 10 K is shown in Figure~\ref{fig:ErXRD}(a) (color online).  Here, K = 3 (remember that $b$ is the long axis) L $\in$ [3.5 5.5] and H $\in$ [0.5 3.5].  Filled symbols are those that were investigated, open symbols are those assumed to be present from symmetry.  Bragg peaks for the average structure are marked by large (black) circles.  Note that for K $>$ 0 these are extinct for H + K = odd. The superlattice peaks $q_1$ $\approx$ 2/7 $c^*$ are shown as (red) upwards-pointing triangles, with second harmonics in smaller (pink) upwards-pointing triangles.  The superlattice peaks $q_2$ $\approx$ 1/3 $a^*$ are shown as (blue) triangles pointing to the right, with smaller (cyan) triangles denoting the second harmonics.  Superlattice peaks at the linear combination $q_1$ + $q_2$, denoted as (purple) stars, were also observed, providing further evidence that these two wavevectors are real and present in the same crystallite.

Raw data taken at 155 K using a Ge(111) crystal analyzer are shown in Fig.~\ref{fig:ErXRD}(b) and (c) for superlattice peaks $q_1$ and $q_2$, respectively.  At this temperature, $q_1$ = 0.298(5) $c^*$ and $q_2$ = 0.344(5) $a^*$.  The FWHM for these peaks is larger for ErTe$_3$ due to a larger crystal mosaic, yet it remains clear that the magnitudes of $q_1$ and $q_2$ are very close but not identical.

Temperature dependent measurements were performed on selected superlattice peaks in ErTe$_3$.  The integrated intensities of a $q_1$ peak (-1 3 2+$q_1$) and a second harmonic of $q_1$ (-1 3 3-2$q_1$) were measured using 1 milliradian slits.  Integrated intensities were normalized to the main Bragg peak (-1 3 2). In Fig.~\ref{fig:ErXRD}(d), as for TbTe$_3$, the square root of the integrated intensity of the first harmonic is plotted with the average of the first four data points (from 25 to 100 K) normalized to one.  This agrees well with the BCS order parameter, which is plotted using a transition temperature of $T_{c1}$ = 267(3) K  as obtained from resistivity, and is shown as a dashed curve.  The second harmonic superlattice peak can be considered as either a second order diffraction harmonic of the first-order superlattice peak, or as the 2$q_1$ component of a non-sinusoidal modulation.\cite{pouget_structural_1985} In the first case, the integrated intensity of the second harmonic would be proportional to the order parameter to the fourth power.\cite{wilson_x-ray_1962}  In Fig.~\ref{fig:ErXRD}(d), the fourth root of the integrated intensity of (-1 3 3-2$q_1$), with the average of first four data points normalized to one, is shown to agree with the BCS curve.  This is consistent with the case of the purely sinusoidal modulation, although the presence of non-sinusoidal components cannot be ruled out.   Just above $T_{c2}$ there is small dip in intensities of both peaks.  Although this could be a sign of the second CDW interacting with the first, errors are such that this dip should not be overly interpreted.  Error bars shown are $\pm$ 10 $\%$ of integrated intensity for both peaks.

Using a Ge(111) crystal analyzer, the temperature dependence of the $q_2$ peak (1-$q_2$ 6 4) was followed from 10 K to 180 K .  Shown in Fig.~\ref{fig:ErXRD}(e) is the integrated intensity from 140 K to 180 K.    The vertical dashed line is $T_{c2}$= 159(5) K as obtained from resistivity measurements.    The integrated intensity decreases rapidly with increasing temperature across $T_{c2}$, similar to the behavior of $q_1$ through $T_{c1}$.  Error bars shown are $\pm$ 20$\%$.  As in the case for TbTe$_3$ (Fig.~\ref{fig:Tbtransition}(a)), scattering intensity is observed for several tens of kelvin above $T_{c2}$, indicative of the presence of substantial fluctuations.

\section{Discussion}

Here we address the magnitude of $T_{c1}$ for these materials,
the origin of the second transition for the heaviest members of the rare earth series,
and the variation in $T_{c1}$ and $T_{c2}$ with chemical pressure.

First we comment on the absolute value of $T_{c1}$.
Taking TbTe$_3$ as an example, if the maximum value of the CDW gap
as measured by ARPES ($\Delta$ $\sim$ 240 meV) \cite{brouet_comm}
is used to estimate a mean-field transition temperature via the familiar BCS expression,
one obtains $T_{MF}(max)$ $\sim$ 1600 K,
over four times greater than the observed $T_{c}$ of 336 K.
However, since the FS of $R$Te$_3$ is not perfectly nested,
the mean field transition temperature will
be substantially reduced from this value. \cite{Yamaji_1982, Yamaji_1983, Maki}
Furthermore, the observation of superlattice peaks with rapidly increasing FHWM
and decreasing correlation length well above $T_{c1}$
show the presence of substantial fluctuations.
These further reduce the critical temperature from the mean field value,
as has been previously observed for similar quasi-1D and 2D materials. \cite{gruner1}

To gain insight to the variation in the observed CDW transition temperatures,
band structure calculations were performed for the unmodulated crystal structure for several members of the rare earth series, using the linear muffin-tin orbital (LMTO) method within the atomic sphere approximation, including combined-correction terms. \cite{barbiellini}  The 4f-electrons of the rare-earth elements
were treated as open-core states.
As input parameters for these calculations, we have used lattice parameters
measured at 300 K. For compounds for which 300 K is above the CDW transition, we have used the refined atomic positions
in the $Cmcm$ space group as published by Malliakas \emph{et al}. \cite{malliakas_2006} For compounds for which 300K is below $T_{c}$ we have used the unmodulated $Cmcm$ space group, but have taken the internal parameter
(which defines the $y$-position of the Te planes in the unit cell)
from atomic refinements to the 4-dimensional $C2cm(00\gamma)000$ space group \cite{malliakas_2006} at 300 K.

\begin{figure}
\subfigure{
    \includegraphics[width =1.6 in]{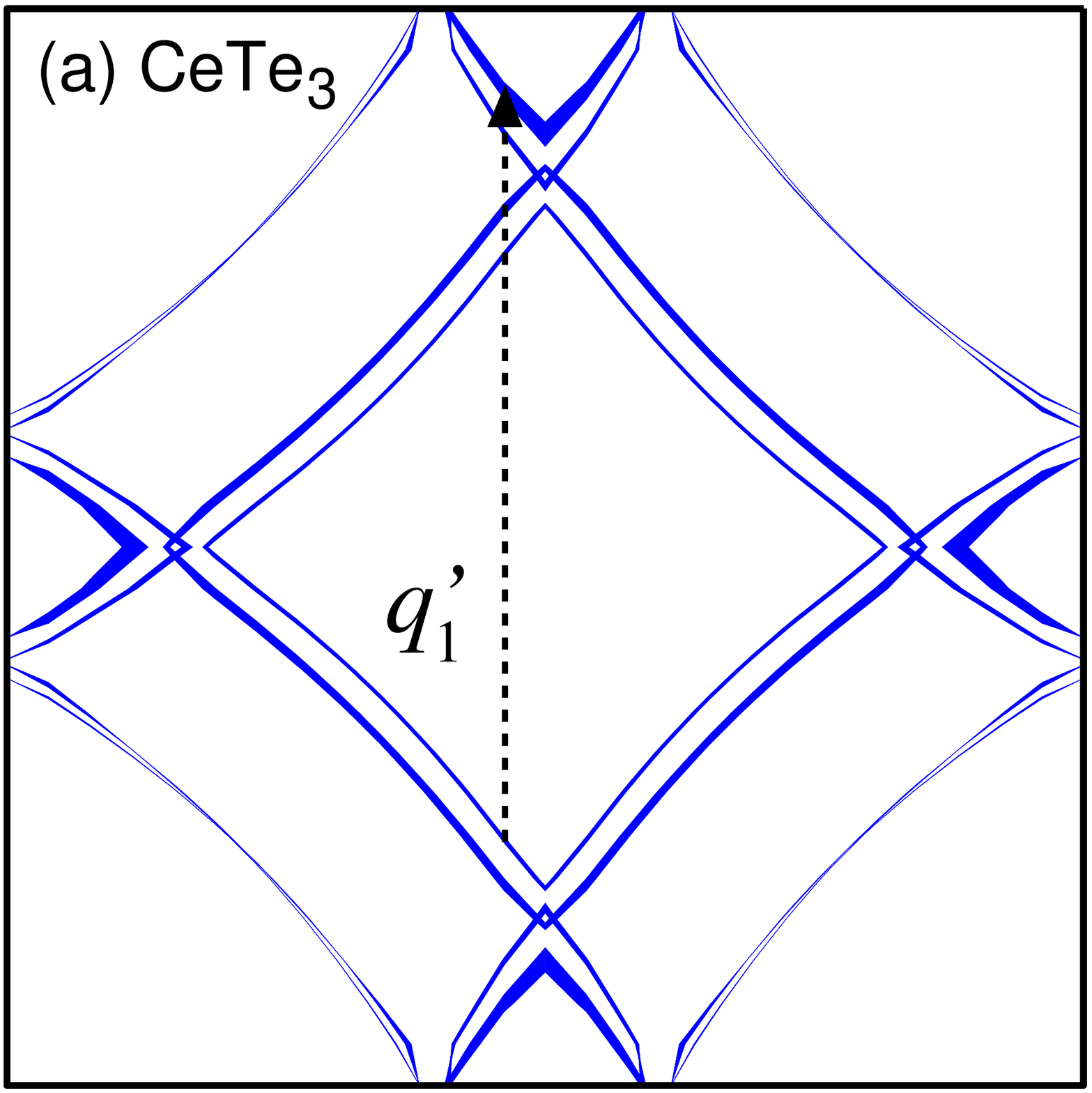}}
\subfigure{
    \includegraphics[width=1.6in]{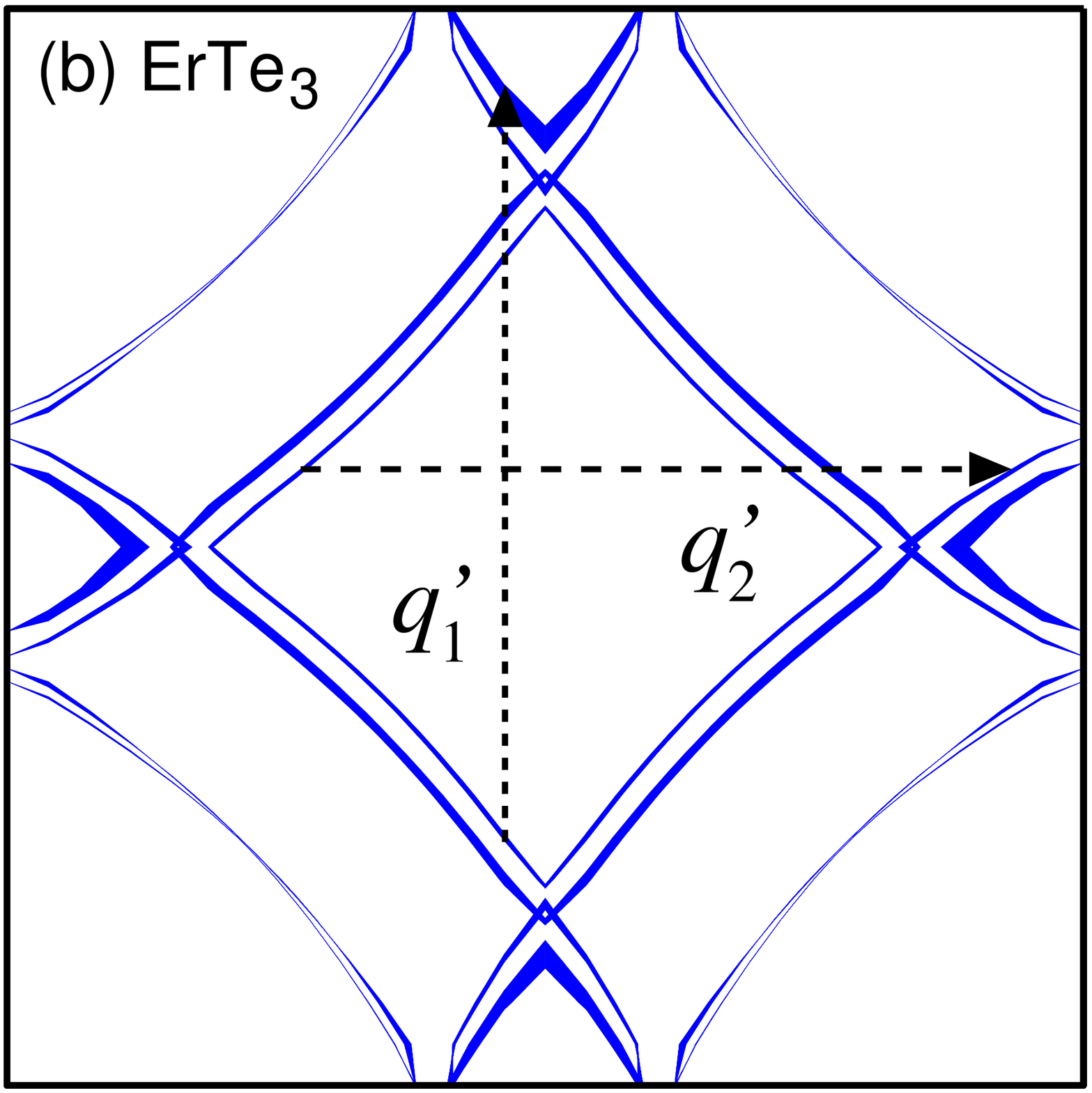}}
\caption{\label{fig:FS}Fermi surface of (a) CeTe$_3$ and (b) ErTe$_3$ obtained from LMTO band structure calculations.  The Fermi surface is shown in projection down the $b^*$-axis such that the thickness of the lines is indicative of the $b^*$-axis
dispersion of each of the bands..  The horizontal dimension is $k_x$ in the interval $\pm$ $a^*$/2.  The vertical dimension is $k_z$ in the interval $\pm$ $c^*$/2.  The wavevectors $q_1'$ and $q_2'$ and corresponding nesting conditions are marked.}
\end{figure}

\begin{figure}[t]
\includegraphics[width=3.5in]{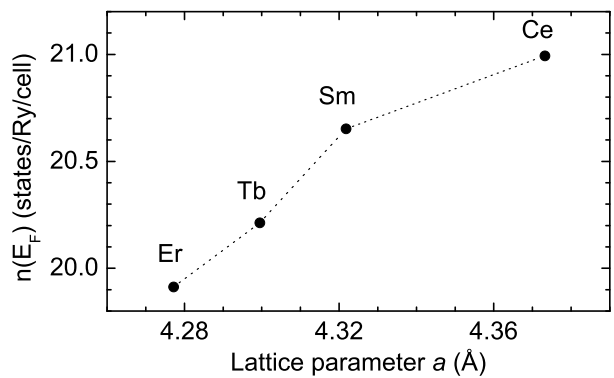}
\caption{\label{fig:DOS} Density of states at $E_F$ for representative members of the $R$Te$_3$ series, as obtained from LMTO band structure calculations, shown as a function of the in-plane lattice parameter $a$.  Dashed lines are drawn between the points to guide the eye.}
\end{figure}

Typical results illustrating the FS seen in projection down the $b^*$-axis are shown in Fig.~\ref{fig:FS} for two specific compounds close to the ends of the rare earth series, CeTe$_3$ and ErTe$_3$. The calculated Fermi surfaces for these two isoelectronic compounds are, as anticipated, broadly similar, with only subtle differences in topology at the very corners of the diamond-like sheets of FS.  Both the bilayer splitting and the dispersion along the $b^*$-axis (i.e. the thickness of the lines in Fig.~\ref{fig:FS}) evolve weakly from CeTe$_3$ to ErTe$_3$, by $\sim$ +4$\%$ in the first case, as measured by the separation of the two inner diamond sheets at $k_x$ = 0.16 $a^*$,  and by $\sim$ +12$\%$ in the latter, also measured at the inner diamond sheets at $k_x$ = 0.16 $a^*$.  The overall bandwidth increases from CeTe$_3$ to ErTe$_3$, and the density of states at $E_F$, $n(E_F)$, decreases almost linearly from 21.1 for CeTe$_3$ to 19.9 states/Ry/cell for ErTe$_3$, a variation of approximately 5$\%$ (Fig.~\ref{fig:DOS}).

Quantitative calculations of the critical temperature are difficult for an imperfectly nested system, but it is clear that all three effects described above will contribute to the decrease in $T_{c1}$ with decreasing lattice parameter. The subtle variation in the $b^*$ axis dispersion and the bilayer splitting result in poorer nesting conditions, while the reduction in $n(E_F)$ directly affects the mean field transition temperature. In the absence of more detailed calculations we cannot determine the exact extent to which each of these effects contributes to the variation in $T_{c1}$ across the rare earth series. However, given that the mean field transition temperature varies exponentially with $n(E_F)$ it is likely that the
variation in $n(E_F)$ plays the dominant role.

The second CDW transition, observed only for the heaviest members of the
rare earth series, is particularly intriguing. Specifically,
in contrast to most usual CDW transitions, the effect of
increasing (chemical) pressure is actually to \textit{increase} $T_{c2}$.
It appears that this unusual behavior is directly linked to the variation in $T_{c1}$.
In particular, the extent to which the FS is gapped below $T_{c1}$ depends on
the maximum size of the CDW gap. A reduction in the maximum gap causes a reduction in the total area of the FS that is
gapped precisely because the FS is imperfectly nested.
This is borne out by recent ARPES experiments, which reveal that for the
compounds with smaller lattice parameters the CDW gap
extends a smaller distance around the FS from the tips of the diamond sheets
at $k_x$ = 0. \cite{maxgap, brouet_comm} Hence, on traversing the rare
earth series from LaTe$_3$ towards TmTe$_3$, progressively more of the FS remains intact below
$T_{c1}$, and as a consequence is available for a subsequent CDW transition at a lower temperature.
Having already lost the ``tips'' of the diamond sections of the FS that point in the
$c$-axis direction at the first transition,\cite{maxgap} the second transition must involve sections
of the FS closer to the tips of the diamonds pointing in the transverse $a$-axis direction.
Indeed, these are exactly the regions of the FS that contribute to the competing peak
in $\chi(q)$ found at ~ 2/7 $a^*$ in band structure calculations for the unmodulated
structure.\cite{chiq} Preliminary ARPES results for ErTe$_3$ confirm this picture, revealing additional gaps
forming on sections of the FS close to the tips of the diamond sections of the FS pointing in the $a^*$ direction.\cite{moore_comm} The corresponding jump in the resistivity at $T_{c2}$, related to the amount of FS gapped
at the transition, is largest for the compound with the largest value of $T_{c2}$ (smallest value of
$T_{c1}$) and smallest area of initial FS gapped at $T_{c1}$. The resulting ``rectangular'' CDW state is characterized by orthogonal, independent wave vectors, almost equal in magnitude, but with an order parameter that is developed to a different
degree in the two directions.

Yao and coworkers have recently addressed the question of the origin of the unidirectional CDW for the lightest members of the $R$Te$_3$ series.\cite{yao} Using a simple tight-binding model for a square lattice, and considering only non-interacting Te $p_x$ and $p_z$ bands, they found that for sufficiently high values of the electron-phonon coupling (and consequently $T_{c}$) a unidirectional (``stripe'') CDW at 45 degrees to the Te-Te bonds is favored over a bidirectional (``checkerboard'') state in which $p_x$ and $p_z$ sheets of the FS are individually gapped by orthogonal wave vectors which are oriented at a different angle to the stripe phase. In this model, the stripe phase gaps a sufficient amount of the FS that only one of the two equivalent directions ($a$ and $c$ in the real material) is chosen for the lattice modulation. The weak orthorhombicity of $R$Te$_3$ provides a natural symmetry-breaking between the $a$ and $c$ axes such that for the lightest members of the series the unidirectional CDW is always observed along the $c$-axis and never along $a$. \cite{chiq} It will be intriguing to see whether this insightful toy model can be extended to incorporate the observed ``rectangular'' CDW phase at a more quantitative level than the qualitative arguments presented above.

Implicit in all of the above discussion has been the assumption that the electron-phonon coupling itself does not have a strong q-dependence and does not vary substantially across the rare earth series, though it is not necessarily immediately
obvious that this should be the case. The Te-Te bond-length in $R$Te$_3$ is substantially longer than the standard covalent
bond-length, implying that the unmodulated high-temperature structure of $R$Te$_3$
is inherently unstable, regardless of the low-dimensional electronic structure.\cite{malliakas_2005}  The close correspondence between the maxima in $\chi(q)$ and the observed superlattice wave vectors clearly demonstrates that the electronic system
plays a preeminent role in the lattice modulation in this material.  But it is not clear whether it is the \textit{only} driving force for the instability.
To determine the relative role played by the lattice will take a substantial computational effort.
However, this material appears to be a model system to quantitatively calculate the role played by an incipient lattice instability in CDW formation, something which to date has been lacking for most other CDW systems. \cite{mazin_comm}

\section{Conclusions}

The $R$Te$_3$ family of compounds have a remarkable phase diagram. All members of the
series exhibit a CDW transition to a state characterized by a simple unidirectional
incommensurate lattice modulation with wave vector $q_1$ $\approx$ 2/7 $c^*$. The transition temperature $T_{c1}$ is very sensitive to chemical pressure, varying by over 200 K across the series, principally ascribed to the variation in the density of states at the Fermi level $n(E_F)$. The heaviest members of the series exhibit an additional CDW transition at a lower
temperature $T_{c2}$, with an almost equivalent wavevector oriented perpendicular to the first, along the $a^*$-axis.
The resulting ``rectangular'' CDW state consists of perpendicular modulation wave vectors
almost equal in magnitude, but with independent order parameters, each developed to a
different degree at any given temperature. These observations can be understood in terms of the second CDW nesting regions of the original FS that are ungapped by the first CDW. Specifically, on traversing the rare earth series, as
$T_{c1}$ decreases, increasingly more of the remaining FS becomes available to drive a second CDW with a modulation wave
vector transverse to the first.

We conclude by pointing out that by preparing samples with a continuous solid solution between adjacent rare earths in the lanthanide series it would  be possible to continuously tune the lattice parameter from that of LaTe$_3$ to that of TmTe$_3$ while retaining the same band filling. This exquisite degree of control of a physical
parameter directly affecting the transition temperature, in conjunction with the remarkable simplicity of the ``stripe'' and ``rectangular'' CDW states, mark the rare earth tritellurides as a model system for studying the effects of imperfect FS nesting on CDW formation.

\section*{Acknowledgments}
We thank V. Brouet, M. D. Johannes, E. A. Kim, S. A. Kivelson, I. I. Mazin, R. Moore, J. A. Robertson, Z.-X. Shen, and H. Yao for numerous helpful conversations.  Z. Madon contributed to crystal growth.
This work is supported by the DOE, Office of Basic Energy Sciences, under contract number DE-AC02-76SF00515.  IRF is also supported by the Hellman Foundation.
Portions of this research were carried out at the Stanford Synchrotron Radiation Laboratory,
a national user facility operated by Stanford University on behalf of the US Department of Energy, Office of Basic Energy Sciences.

%\bibliography{Ru_CDWtransition_longpaper_submitted}

\begin{thebibliography}{10}
\bibitem{maxgap}
ARPES measurements for CeTe$_3$ reveal a maximum gap of 400 meV at the Fermi
  energy for $k_x$ = 0 (i.e. the ``tips'' of the diamond sheets of the FS that
  are pointing along the $c^*$ direction - see Fig 8). The gap diminishes in
  magnitude the further one goes around the FS from this point, reaching zero
  at approximately $k_x$ = 0.5 $a^*$.\cite{brouet} In comparison, for SmTe$_3$,
  for which the maximum gap is only 280 meV, the gap extends only as far as
  $k_x$ = 0.35 $a^*$ \cite{brouet_comm}.
\bibitem{TbTe3_surface}
While a second lower transition was not observed in the resistivity for
  TbTe$_3$, STM measurements on TbTe$_3$ at 6K \cite{fang} show an ordering
  with wavector $\sim$ 2/3 $a^*$ which is similar to $q_2'$ a observed by x-ray
  diffraction in ErTe$_3$ below $T_{c2}$. This could mark the appearance of an
  extraordinary transition present only on the surface, or it could indicate
  that TbTe$_3$ does in fact have a second transition with $T_{c2}$ $>$ 6 K,
  but with accompanying resistivity feature too small to be observed.
\end{thebibliography}
%\bibliographystyle{prsty}

 \end{document}